\documentclass[12pt, draftclsnofoot, onecolumn]{IEEEtran}
\usepackage{algorithm, algorithmic}
\usepackage{amsmath, amssymb, scalerel, amsthm, cite, relsize}
\newtheorem{proposition}{Proposition}
\newtheorem{theorem}{Theorem}
\newtheorem{example}{Example}

\newcommand{\pgt}{P_{\scaleto{\mathrm{GT}}{3.5pt}}}
\newcommand{\pgtd}{P_{\scaleto{\mathrm{GT}}{3.5pt},{\scaleto{\mathrm{D}}{3.5pt}}}}
\newcommand{\puav}{P_{\scaleto{\mathrm{UAV}}{3.5pt}}}
\newcommand{\pdist}{P_{\scaleto{\mathrm{D}}{3.5pt}}}

\title{Power-Efficient Deployment of UAVs as Relays}
\author{Erdem Koyuncu\thanks {The author is with the   Department of Electrical and Computer Engineering, University of Illinois at Chicago. Email: ekoyuncu@uic.edu.
    }}
    
\begin{document}
\maketitle
\begin{abstract}
Optimal deployment of unmanned aerial vehicles (UAVs) as communication relays is studied for fixed-rate variable-power systems. The considered setup is a set of ground transmitters (GTs) wishing to communicate with a set of ground receivers (GRs) through the UAVs. Each GT-GR pair communicates through only one selected UAV and have no direct link. Two different UAV selection scenarios are studied: In centralized selection, a decision center assigns an optimal UAV depending on the locations of all terminals. In distributed selection, a GT selects its relaying UAV using only the local knowledge of its distances to the UAVs. For both selection scenarios, the optimal tradeoff between the UAV and GT power consumptions are determined using tools from quantization theory. Specifically, the two extremal regimes of one UAV and very large number of UAVs are analyzed for a path loss exponent of $2$. Numerical optimization of UAV locations are also discussed. Simulations are provided to confirm the analytical findings.
\end{abstract}
\section{Introduction}
Unmanned aerial vehicles (UAVs) and their networks are becoming increasingly relevant in many practical applications in a variety of communication and information systems \cite{tutopap}. In the context of wireless communications, UAVs can be used to establish a temporary communication infrastructure during a natural disaster \cite{disaster}, or to provide coverage to remote areas by acting as drone cells \cite{yanikomertut}, among many other use cases \cite{tutopap}.

Power-efficiency is a fundamental issue in the design of any wireless network. Especially in the case of wireless system that incorporates UAVs, power-efficiency is a key design goal as the act of flight typically needs a significant amount of power. The availability of high mobility, on the other hand, provides the opportunity of placing the UAVs with respect to the networking environment. A fundamental problem is then to determine, given networking conditions, the optimal UAV deployment that results in the most power-efficient operation.

There are several studies on the placement and/or trajectory optimization of UAVs for different objectives. Several works, including \cite{bor2016efficient, lyu1, mozaffari1, shaka1}, have considered the static placement of UAVs as mobile base stations to maximize network coverage. In particular, \cite{mozaffari1, shaka1} optimize the UAV deployment to minimize the average UAV transmission power subject to ground terminal coverage or rate constraints. For time-varying networks, trajectory optimization of UAVs serving one or more ground terminals have also been extensively studied \cite{zeng2017energy, jiang2012optimization, marier1, koyuncu1}. In \cite{aviationtime}, the authors present algorithms and analytical solutions for minimizing the flight time of UAVs that are employed for sensor data collection.

The idea of cooperation and/or relaying to enhance the performance of wireless systems goes back to many decades ago and has  been the subject of many publications \cite{nosra, koyuncu2}. Recently, there has also been a lot of work on relaying in the context of UAVs. In particular, a mobile relaying method for a single source-destination pair and one UAV has been developed in \cite{zeng2016throughput}. In \cite{zhan2011wireless}, the authors study the case of several UAVs serving as relays between several ground terminals and a base station, and develop a joint heading and adaptive handoff algorithm.
The deployment of UAVs for intercellular traffic offloading has been studied in \cite{rohde1}. A stochastic programming approach to UAV relay trajectory optimization can be found in \cite{athina}. In \cite{shuhang1}, the authors study the UAV transmit power optimization problem for an amplify-and-forward relay acting in between a single user and a base station. Field test results on the performance of relaying UAVs are also available \cite{weisi1}. 
 
Most of the above work on relaying UAVs consider a numerical approach for finding the optimal UAV deployments and the corresponding network performance. Some only consider the case of a single UAV. In this work, we consider power-efficient deployment of any number of UAVs as relays between ground terminals (GTs) and ground receivers (GRs) of arbitrary densities. We analytically characterize the minimum GT (signal) transmission power subject to a maximum UAV transmission power and vice versa. We also find the corresponding optimal deployment of UAVs. Our approach relies on tools from quantization theory \cite{grayneuhoff}. Another work \cite{koyuncu1} also utilizes quantization theory to analyze networks of UAVs acting as mobile base stations. The results of \cite{koyuncu1} are, however, not applicable to relaying UAVs due to the fundamentally different problem formulations and objective functions. We refer to \cite{VD, hjdeploy, koyuncuxx0, koyuncuxx1} for applications of quantization methods to other problems that involve deployment of wireless nodes.


The rest of this paper is organized as follows: In Section \ref{secSystemModel}, we introduce the system model. We analyze the cases of centralized and distributed UAV selection strategies in Sections \ref{secCentralized} and \ref{secDistributed}, respectively. We provide numerical simulation results in Section \ref{secNumerical}. Finally, in Section \ref{secConclusions}, we draw our main conclusions and discuss several directions for further research.

\section{System Model}
\label{secSystemModel}
We consider several GTs and GRs at zero
elevation and several UAVs at a fixed elevation $h > 0$.
Mathematically, we assume that the GTs and GRs are located on $\mathbb{R}^d$, where $d \in \{1,2\}$, and the UAVs are located on  $\mathbb{R}^d \times \{h\}$, where $\times$ is the Cartesian product. The case $d=1$ is relevant when the GTs and GRs are located on a straight line, e.g. on a highway. 

We model the location of a given GT as a random variable $X$ with probability density function $f_X$. Likewise, we model the location of a given GR as a random variable $Y$ with density $f_Y$. In practice, the GTs may correspond to a network of sensors, which transmit sensed data unidirectionally to a collection of access points (GRs) through the UAVs. 

Let $u_1, \ldots, u_n$ denote the projected locations of the $n$ UAVs to $\mathbb{R}^d$. We assume that there is no direct link available between the GTs and the GRs due to geographical obstructions. The message of a given GT is first transmitted to one of the UAVs, which forwards the GT message to the target GR. More specifically, we consider a rate-$\rho$ variable-power transmission scheme where a GT at location $x$ wishes to communicate with a GR at location $y$ through the $i$th UAV at location $u_i$. Due to the aerial nature of the communication system, we assume that the multipath effects are negligible and the transmitted signals only undergo path loss with exponent $r$. In such a scenario, the channel capacity between the GT at $x$ and the UAV at $u_i$ is given by $\smash{\log_2(1 + \frac{P_0}{d(x, u_i)})}$ bits/sec/Hz, where
\begin{align}
d(x, u_i) \triangleq  (h^2 + \|x-u_i\|^2 )^{\frac{r}{2}}, 
\end{align}
and $P_0$ represents the transmission power of the GT in Joules per second. Similarly, the channel capacity between the UAV at $u_i$ and the GR at $y$ is given by $\smash{\log_2(1 + \frac{P_0}{d(u_i, y)})}$, where $P_1$ is the transmission power of UAV $i$. The GT may successfully communicate with the GR at rate $\rho$ if and only if the channel capacities of the GT to UAV $i$ link and the UAV $i$ to GR link are both no less than $\rho$. In other words, the conditions $\smash{\log_2(1 + \frac{P_0}{d(x, u_i)})} \geq \rho$ and $\smash{\log_2(1 + \frac{P_1}{d(u_i, y)})} \geq \rho$
should both be satisfied. These are equivalent to the power constraints $P_0\geq (2^{\rho} - 1)d(x,u_i)$ and $P_1\geq (2^{\rho} - 1)d(u_i,y)$. Setting $\rho = 1$ without loss of generality, the minimum GT and UAV transmission powers that guarantee reliable communications are then given by $d(x, u_i)$ and $d(u_i, y)$, respectively.

In general, the relaying UAV index may depend on both the GT and GR locations. More precisely, given the GT location $x$ and the GR location $y$, suppose that the relaying UAV index is given by $I(x,y)$. The (minimum) average GT transmission power given the UAV deployment $U =  (u_1, . . . , u_n)$ and the UAV index assignment function $I$ is then given by
\begin{align}
\label{pgui}
\pgt(U,I) \triangleq \iint d(x,u_{I(x,y)})f_{X,Y}(x,y)\mathrm{d}x\mathrm{d}y,
\end{align}
where $f_{X,Y}(x,y) = f_X(x)f_Y(y)$ is the joint density function of the GTs and GRs. Similarly, the average UAV transmission power for reliable communication can be expressed as 
\begin{align}
\label{paui}
 \puav(U,I) \triangleq \iint d(u_{I(x,y)},y)f_{X,Y}(x,y)\mathrm{d}x\mathrm{d}y.
\end{align}

On one extreme, the UAVs may be positioned closer to
where the GTs are densely populated, resulting in low GT
transmission power. This may come at the expense of possibly
high UAV transmission power. On the other extreme, the
UAVs may be positioned closer to GRs, resulting in low
UAV transmission power. Our general goal is  to obtain
the tradeoff 
\begin{align}
\label{pgstar}
\pgt^{\star}(p) \triangleq \min\{\pgt(U,I): \puav(U,I) \leq p\}
\end{align}
for different $p$. The minimization is over all deployments $U$ and index assignments $I$. An equivalent
problem is to instead minimize $\puav(U,I)$ subject to a constraint
on $\pgt(U,I)$.

An inherent assumption of our formulation so far is the
availability of a decision center that can always select the best
UAV given GT, GR and UAV locations. We call this scenario
the centralized UAV selection scenario, which amounts to allowing optimization over
all possible index assignment functions $I$, as in (\ref{pgstar}). We shall
later also study distributed selection schemes that use only
local knowledge at the GTs or UAVs. To study such schemes, we will impose restrictions
on the possible index assignment functions we can consider.

\section{Centralized UAV Selection}
\label{secCentralized} 
We first study the case of centralized UAV selection, where
we are concerned with the calculation of (\ref{pgstar}). Let
\begin{align}
\label{lagrangian1}
P(U, I) \triangleq \pgt(U, I) + \lambda \puav (U, I). 
\end{align}
Minimizing $P(U, I)$ over $U,I$ for different values of
the Lagrange multiplier $\lambda$ allows travel over the $(p, \pgt^{\star}(p))$
tradeoff curve.\footnote{More precisely, this provides travel over only the convex hull of the
tradeoff curve. Nevertheless, in practice, given UAV-GT transmission pairs
$(\puav^1, \pgt^1)$ and $(\puav^2, \pgt^2)$, the pair $(\tau \puav^1 + (1-\tau) \puav^2, \tau \pgt^1 + (1-\tau) \pgt^2)$
is achievable for any $0 \leq \tau \leq 1$ by simply time sharing between the two
corresponding UAV deployments. As a result, our Lagrangian approach can
be considered to yield, in fact, the entire ``operational'' tradeoff curve, which
is the convex hull of the ``actual'' tradeoff curve $(p, \pgt^{\star}(p))$. For a simpler exposition, we shall assume throughout the paper that $(p, \pgt^{\star}(p))$ is convex, in which case the operational curve coincides with the actual curve and no distinction between the two is necessary. } For example, a small $\lambda$ does not penalize the
UAV transmit power consumption as compared to a larger $\lambda$.
It thus results in a lower UAV power consumption compared
to the case of a larger $\lambda$, albeit at the expense of a larger GT
power consumption. The Lagrangian approach to entropy-
constrained vector quantizers \cite{ecvq} follows a similar
formulation.

Substituting (\ref{pgui}) and (\ref{paui}) to  (\ref{lagrangian1}), we have
\begin{align}
\label{lagrangian2}
 P(U,I) = \iint [ d(x,u_{\scaleto{I(x,y)}{6pt}})+ \lambda d(u_{\scaleto{I(x,y)}{6pt}},y) ] f_{X,Y}(x,y)\mathrm{d}x\mathrm{d}y,
\end{align}
The optimal
index mapping that minimizes (\ref{lagrangian2}) is thus
\begin{align}
\label{aystarxy}
I^{\star}(x,y) \triangleq \arg\min_{1 \leq i \leq n}\bigl[ d(x,u_i)+ \lambda d(u_i,y) \bigr].
\end{align}
Minimizing (\ref{lagrangian2}) over all $U, I$ is then equivalent to minimizing
\begin{align}
\label{lagrangian3}
 P(U) \triangleq \iint \min_i \bigl[ d(x,u_i)+ \lambda d(u_i,y) \bigr] f_{X,Y}(x,y)\mathrm{d}x\mathrm{d}y
\end{align}
over all deployments U.
\subsection{The Case $r=2$}
We first consider the special case where the path loss exponent
is given by $r=2$. In addition to its practical significance, the case  $r=2$ allows a precise theoretical characterization of achievable
performance as we shall demonstrate. Without loss of generality, we assume $h=0$ when $r=2$, as the effect of a non-zero height $h$ when $r=2$ is simply an extra additive power consumption of $h^2$ for both UAVs and GTs. Before proceeding further, let us also define the constants 
\begin{align}
c_0 & \triangleq \mathrm{E}\|X - Y\|^2, \\
c_1 & \triangleq \|\mathrm{E}X-\mathrm{E}Y]\|^2, \\
c_2 &  \triangleq \mathrm{E}\|X - \mathrm{E}Y\|^2, \\ 
c_X & =  \mathrm{E}\|X\|^2 - \|\mathrm{E}X\|^2, \\
c_Y & =  \mathrm{E}\|Y\|^2 - \|\mathrm{E}Y\|^2
\end{align}
that we will frequently use. Here, ``$\mathrm{E}$'' is the expected value.

Given $r=2$, the integrand in (\ref{lagrangian3}) can be rewritten
as
\begin{align}
   \label{skajdkasjda1} d(x,u_i) +\lambda d(u_i,y)  
 &  = \|x-u_i\|^2 + \lambda \|y - u_i\|^2 \\
\label{skajdkasjda2}  &    = (1+\lambda) \left\| u_i - \frac{x+\lambda y}{1+\lambda} \right\|^2 + \frac{\lambda\|x - y\|^2}{1+\lambda}.
\end{align}
The final equality (\ref{skajdkasjda2}) can simply be verified by expansion
of the squared Euclidean norms in (\ref{skajdkasjda1}) and (\ref{skajdkasjda2}). Letting
\begin{align}
 Z \triangleq \frac{X+\lambda Y}{1+\lambda},
\end{align}
and substituting (\ref{skajdkasjda2}) to (\ref{lagrangian3}), we obtain
\begin{align}
\label{lagrangian4}
 P(U) = \frac{\lambda c_0}{1+\lambda} + (1+\lambda) \int \min_{1\leq i \leq n} \|u_i - z\|^2 f_Z(z)\mathrm{d}z.
\end{align}
The second term is the distortion
of an optimal quantizer given reproduction points $u_1,\ldots,u_n$,
scaled by the constant $(1 + \lambda)$. Therefore, finding an optimal UAV relay deployment is
equivalent to designing an optimal quantizer for a linear
combination $Z$ of the ground transmitter and receiver densities. Existing results and tools from quantization theory can be applied to solve the UAV relay deployment problem. Nevertheless, an exact solution to the minimization of $P(U)$ is only available for a few special cases. One special case is that of a single UAV $n=1$, where the optimal deployment that minimizes (\ref{lagrangian4}) is well-known to be $u_1 = E[Z]$. Substituting to (\ref{pgui}) and (\ref{paui}) leads to the following after some straightforward calculations.
\begin{proposition}
\label{oneuavprop}
Let $n=1$, $r=2$. The UAV-GT transmission power tradeoff $(p, \pgt^{\star}(p))$ can be parameterized as
\begin{align}
\left( c_Y+\frac{c_1}{(1+\lambda)^2},c_X+ \frac{c_1 \lambda^2}{(1+\lambda)^2}\right),\,\lambda \geq 0. 
\end{align}
The optimal UAV location is 
\begin{align}
u_1^{\star} = \frac{\mathrm{E}[X]+\lambda \mathrm{E}[Y]}{1+\lambda}. 
\end{align}
In particular, given a UAV power constraint $p\in [c_Y, c_Y + c_1]$, the minimum GT power consumption can be found by setting $\lambda = \sqrt{\vphantom{X_{x_x}}\smash{\frac{c_1}{p-c_Y}}} - 1$. This yields
\begin{align} 
\!\pgt^{\star}(p) =  c_X + \left(\sqrt{c_1} - \sqrt{p-c_Y}\right)^2,\,p\in [c_Y, c_Y + c_1].\!
\end{align}
\end{proposition}
\begin{example} 
\label{example1}
Let  $f_X(x) = \mathbf{1}(x\in[0,1])$ and $f_Y(y) = \mathbf{1}(y\in[2,3])$, where $\mathbf{1}(\cdot)$ is the indicator function. We thus consider uniform densities on two non-intersecting unit intervals that are within unit distance. By Proposition \ref{oneuavprop}, given $\lambda \geq 0$, the optimal UAV location is $E[Z] =\frac{1 + 5\lambda}{2(1+\lambda)}$, resulting in the average GT and UAV power consumptions 
\begin{align}
p_{\lambda} & \triangleq \frac{1}{12}+\frac{4}{(1+\lambda)^2}, 
\end{align}
and
\begin{align}
\pgt^{\star}(p_{\lambda}) & = \frac{1}{12}+\frac{4\lambda^2}{(1+\lambda)^2},
\end{align}
respectively. Equivalently, we have the tradeoff curve
\begin{align}
\pgt^{\star}(p) = \frac{1}{12} + \left(2  - \sqrt{p-\frac{1}{12}}\right)^2,\,p\in\left[\frac{1}{12},\frac{49}{12}\right].
\end{align}
This concludes our example. {\hfill\ensuremath{\square}}
\end{example}

We now consider a general $n>1$. For an origin-centered set $A$ (i.e., $\int_A x \mathrm{d}x = 0$), let
\begin{align}
\kappa(A) \triangleq \frac{\int_A \|x\|^r \mathrm{d}x}{(\int_A \mathrm{d}x)^{\frac{d+r}{d}}} 
\end{align}
denote the normalized $r$th moment of $A$. Regarding the minimization of (\ref{lagrangian4}), a fundamental result in quantization theory is the existence of constants $\kappa_{rd}>0$ such that \cite{bennett1, zador1}
\begin{align}
\label{optdeployment}
 \min_U \int \min_{1\leq i \leq n} \|u_i - z\|^r f_Z(z)\mathrm{d}z = \kappa_{rd}n^{-\frac{r}{d}} \|f_Z\|_{\frac{d}{d+r}} + o(n^{-\frac{r}{d}})
\end{align}
as $n\rightarrow\infty$. Here, 
\begin{align}
\|f_Z\|_p & \triangleq \left(\int (f_Z(z))^p \mathrm{d}z\right)^{\frac{1}{p}}
\end{align}
is the $p$-norm of $f_Z$. In particular, it can be shown that $\kappa_{r1}$ and $\kappa_{r2}$ are given by the normalized $r$th moments of the origin-centered interval and the hexagon, respectively. For example, $\kappa_{21} = \frac{1}{12}$ and $\kappa_{22} = \frac{5}{18\sqrt{3}}$. A byproduct of the result is that for a uniform distribution and one dimension, the UAVs should be placed in an equispaced fashion. In two dimensions, the optimal arrangement becomes instead a hexagonal lattice.

The optimal UAV deployments that achieve (\ref{optdeployment}) have the following property \cite{grayneuhoff}: Given $Z$, there exists a continuous ``point density function'' $\ell_Z$ such that for any $z$ and infinitesimal $\mathrm{d}z$, the fraction of points in an optimal $U$ that remain on the cube $[z,z+\mathrm{d}z]$ is given by $\ell_Z(z)\mathrm{d}z$. Specifically, 
\begin{align}
\label{pointdenfunc}
 \ell_Z(z) \triangleq \frac{(f_Z(z))^{\frac{d}{d+r}}}{\int (f_Z(z))^{\frac{d}{d+r}} \mathrm{d}z}.
\end{align}
The (optimal) point density function in (\ref{pointdenfunc}) describes, for every point $z$ on the area of interest, the fraction of the $n$ total UAVs that should be located in the immediate neighborhood of $z$. Since, for any $z$, the density $f_Z$ is approximately uniform on $[z+\mathrm{d}z]$, the UAVs on $[z,z+\mathrm{d}z]$ follows the same structure as stated for uniform distributions before: They are equispaced when $d=1$, and lie on a hexagonal lattice when $d=2$.

We can thus obtain the best UAV deployment under the assumption of a large number of UAVs $n\rightarrow\infty$. The corresponding best possible Lagrangian cost can be obtained by substituting (\ref{optdeployment}) to (\ref{lagrangian4}). One complication is that this approach does not immediately lead to any expressions for the GT and UAV power consumptions. In detail, we can so far obtain an optimal UAV deployment, say $U^{\star} = \{u_1^{\star},\ldots,u_n^{\star}\}$ that minimize (\ref{lagrangian3}), at least in the $n\rightarrow\infty$ regime. We now wish to recover the function (cf. (\ref{pgui}))
\begin{align}
\pgt(U^{\star},I^{\star}) =  \iint d(x,u_{I^{\star}(x,y)}^{\star}) f_{X,Y}(x,y)\mathrm{d}x\mathrm{d}y,
\end{align}
and the analogous $\puav(U^{\star},I^{\star})$ that together satisfy
\begin{align}
\pgt(U^{\star},I^{\star})  + \lambda \puav(U^{\star},I^{\star})  = P(U^{\star}). 
\end{align}
To accomplish this, we let 
\begin{align}
V_i^{\star} \triangleq \bigl\{ z\in\mathbb{R}^d: \|z - u_i^{\star}\| \leq \|z - u_j^{\star}\|,\,\forall j \bigr\}
\end{align}
denote the Voronoi cell for UAV $i$ in an optimal deployment. Given $z\in V_i^{\star}$, the chosen UAV is at $u_i^{\star}$, and the conditional average power consumed by the GTs can thus be expressed as $\int f_{X|Z}(x|z) d(x,u_i^{\star})\mathrm{d}x$. Averaging out $Z$, we obtain 
\begin{align}
   \pgt(U^{\star}  ,I^{\star})  =  \sum_{i=1}^n  \int_{V_i^{\star}}  f_Z(z)  \int  f_{X|Z}(x|z) d(x,u_i^{\star})  \mathrm{d}x \mathrm{d}z.  
\end{align}
Finally, following the well-known approximation $z\in V_i^{\star} \implies z = u_i^{\star} + o(1)$ from high resolution quantization theory \cite[Section IV]{grayneuhoff}, we have $d(x,u_i^{\star}) = d(x,z) + o(1)$. This yields, simply,
\begin{align}
\pgt(U^{\star}  ,I^{\star})  &  =  \mathrm{E}[d(X,Z)] + o(1) \\
\label{qwjepqwjepqwe1} &
   =  \frac{c_0\lambda^2}{(1+\lambda)^2}+ o(1) .
\end{align}
A similar argument for $\puav(U^{\star},I^{\star})$ leads to
\begin{align}
\label{qwjepqwjepqwe2}
 \puav(U^{\star},I^{\star}) = \frac{c_0}{(1+\lambda)^2} + o(1).
\end{align}
We summarize the results so far via the following theorem.
\begin{theorem}
\label{theorem1}
For a given Lagrange multiplier $\lambda$, as the number of UAVs $n\rightarrow\infty$, the optimal deployment $U^{\star}$ of UAVs follows the point density function $\ell_Z$ in (\ref{pointdenfunc}). The corresponding GT and UAV power consumptions are given by (\ref{qwjepqwjepqwe1}) and (\ref{qwjepqwjepqwe2}), respectively. In particular, 
\begin{align}
\pgt^{\star}(p)= \left(\sqrt{c_0} - \sqrt{p}\right)^2,\,p\in[0,c_0].
\end{align}
\end{theorem}

We have thus precisely characterized the achievable performance for the case of $n=1$ and $n\rightarrow\infty$ when $r=2$. As we discuss in the following, the remaining range of values of $n$ and the case of a general $r$ can be addressed numerically.

\subsection{The Case of a General $r$}
We now consider the achievable performance for a general $r$ not necessarily equal to $2$. The main difficulty for $r \neq 2$ is that an algebraic manipulation of the form (\ref{skajdkasjda2}) is not available. We thus follow a numerical approach that is based on the Lloyd algorithm \cite{lloyd1, lbg}. The algorithm originates from rewriting (\ref{lagrangian3}) as
\begin{align}
 P(U) \triangleq \sum_{i=1}^n \phi(u_i, {V}_i(U)),
\end{align}
where for any $u\in\mathbb{R}^d$ and ${V}\subset\mathbb{R}^{2d}$, we define
\begin{align}
 \phi(u,{V}) \triangleq  \int_{{V}} \bigl( d(x,u) + \lambda d(u,y) \bigr) f_{\scaleto{\left[\begin{smallmatrix}X \\ Y \end{smallmatrix}
\right]}{8pt}}(\left[\hspace{-1pt}\begin{smallmatrix}x \\ y \end{smallmatrix}\hspace{-1pt}\right])
\mathrm{d}\left[\hspace{-1pt}\begin{smallmatrix}x \\ y \end{smallmatrix}\hspace{-1pt}\right],
\end{align}
and 
\begin{align}
{{V}}_i(U) \triangleq \{[\begin{smallmatrix}x \\ y \end{smallmatrix}
]: d(x,u_i)+ \lambda d(u_i,y) \leq d(x,u_j)+ \lambda d(u_j,y),\forall j \}
\end{align}
is the Voronoi cell for UAV $i$. The algorithm is typically initialized with a random $U$, and then alternates between the two following steps: (1) The Voronoi regions ${{V}}_i(U),i=1,\ldots,n$ are then computed. (2) For each $i\in\{1,\ldots,n\}$, the function $\phi(u_i, {{V}}_i(U))$ is minimized over $u_i$, while keeping ${{V}}_i(U)$ fixed. Since each step results in a lower $P(U)$, the algorithm is guaranteed to converge in a cost function sense.

One complication in implementing the algorithm may be the task of minimizing $\phi(u,{V})$ for a fixed ${V}$. Fortunately, $\phi(u,{V})$ can easily be shown to be convex in $u$, and thus the minimization can be performed in a computationally-efficient manner. In particular, for $r=2$, we have the closed-form solution $\arg\min \phi(u, {V}) = \int_{{V}} \frac{x + \lambda y}{1+\lambda} \mathrm{d}\left[\hspace{-1pt}\begin{smallmatrix}x \\ y \end{smallmatrix}\hspace{-1pt}\right]$.

\section{Distributed UAV Selection}
\label{secDistributed}
The optimal index assignment function in (\ref{aystarxy}) requires the knowledge of both GT and GR locations and all UAV locations at a centralized decision center. In this section, we consider a more practical scenario where the selected UAV index depends only on the GT and UAV locations and can thus be determined locally at the transmitting GT. Such a restriction can be made by considering index assignment functions of the form $I(x,y) = I_{\scaleto{\mathrm{D}}{3.5pt}}(x)$ that depend only on the GT location $x$. The subscript ``$\mathrm{D}$'' stands for ``distributed.'' In this case, the optimal index assignment that minimizes (\ref{lagrangian2}) is given by
\begin{align}
 I_{\scaleto{\mathrm{D}}{3.5pt}}^{\star}(x) \triangleq \arg\min_i \left[ d(x,u_i) + \lambda \int d(u_i, y)f_Y(y)\mathrm{d}y \right].
\end{align}
We wish to minimize the corresponding  cost
\begin{align}
\label{pdistu}
  \pdist(U)  \triangleq  \int \min_i\left[ \hspace{-1pt}d(x,\hspace{-1pt}u_i) + \lambda \int d(u_i,\hspace{-1pt} y)f_Y\hspace{-1pt}(\hspace{-1pt}y\hspace{-1pt})\mathrm{d}y \right]f_X\hspace{-1pt}(\hspace{-1pt}x\hspace{-1pt})\mathrm{d}x.
\end{align}
Similar to the centralized scenario, the case $r=2$ leads to closed-form solutions. In particular, for the case of a single UAV $n=1$, the centralized and distributed selection scenarios are the same and the existing solution in Proposition \ref{oneuavprop} of Section \ref{secCentralized} can be used. For a general $n \geq 1$, after some algebraic manipulations and evaluations over (\ref{pdistu}), we obtain
\begin{align}
\label{distolagrange}
  \pdist(U) = \frac{c_0\lambda+c_Y \lambda^2 }{1+\lambda}+  (1+\lambda) \int \min_i d(u_i,w) f_W(w)\mathrm{d}w,
\end{align}
where $W \triangleq \frac{X+\lambda \mathrm{E}[Y]}{1+\lambda}$. Apart from the constants, (\ref{distolagrange}) is in the same form as (\ref{lagrangian4}), and can thus be minimized using the methods of Section \ref{secCentralized}. In particular, the point density function corresponding to an optimal deployment $U_{\scaleto{\mathrm{D}}{3.5pt}}^{\star}$ of UAVs is given by $\ell_W$, as defined in (\ref{pointdenfunc}).

Also, setting $Z = W$ in (\ref{qwjepqwjepqwe1}) and (\ref{qwjepqwjepqwe2}), respectively, we obtain (after some algebra)
 \begin{align}
\label{qpwoepqmndndndn1}
\pgt(U_{\scaleto{\mathrm{D}}{3.5pt}}^{\star},I_{\scaleto{\mathrm{D}}{3.5pt}}^{\star}) 
& = \frac{c_2 \lambda^2}{(1+\lambda)^2} + o(1), \\
\label{qpwoepqmndndndn2}
\puav(U_{\scaleto{\mathrm{D}}{3.5pt}}^{\star},I_{\scaleto{\mathrm{D}}{3.5pt}}^{\star}) & = c_Y + \frac{ c_2 }{(1+\lambda)^2}+ o(1).
\end{align}
These lead to the following theorem. Let $\pgtd^{\star}(p)$ be the analogue of $\pgt^{\star}(p)$ in (\ref{pgstar}) for the distributed selection scenario.

\begin{theorem}
\label{theorem2}
For a given Lagrange multiplier $\lambda$, as the number of UAVs $n\rightarrow\infty$, the optimal deployment $U_{\scaleto{\mathrm{D}}{3.5pt}}^{\star}$ of UAVs for the distributed selection scheme follows the point density function $\ell_W$. The corresponding GT and UAV power consumptions are given by (\ref{qpwoepqmndndndn1}) and (\ref{qpwoepqmndndndn2}), respectively. In particular, 
\begin{align}
\pgtd^{\star}(p)= \left(\sqrt{c_2} - \sqrt{p - c_Y}\right)^2,\,p\in[c_Y,c_Y+c_2]. 
\end{align}
\end{theorem}
\begin{example} 
\label{example2}
For the same setup as in Example \ref{example1}, by Theorems \ref{theorem1} and \ref{theorem2}, the tradeoffs for the centralized and distributed selection scenarios are 
\begin{align}
\pgt^{\star}(p) = \left(\sqrt{\frac{50}{12}} - \sqrt{p}\right)^2,\,p\in\left[0,\frac{50}{12}\right], 
\end{align}
and 
\begin{align}
\pgtd^{\star}(p) = \left(\sqrt{\frac{49}{12}} - \sqrt{p-\frac{1}{12}}\right)^2,\,p\in\left[\frac{1}{12},\frac{50}{12}\right], 
\end{align}
respectively. {\hfill\ensuremath{\square}}
\end{example}
The advantage of the distributed selection over its centralized version is the lesser amount of required location information at the GTs. In particular, a GT is not required to know the location of the GR it wishes to communicate. The price to pay is the higher UAV power consumption for the same amount of GT power consumption. This is because, in the distributed scheme, the selected UAV (which is the closest UAV to the GT) may be very far to the target GR in certain cases. This results in very high UAV power consumption to compensate for the long distance communication. The centralized scheme follows a more balanced approach by taking into account the required power for both hops of communication.

\begin{figure}[h]
\begin{center}
\scalebox{0.46}{\includegraphics{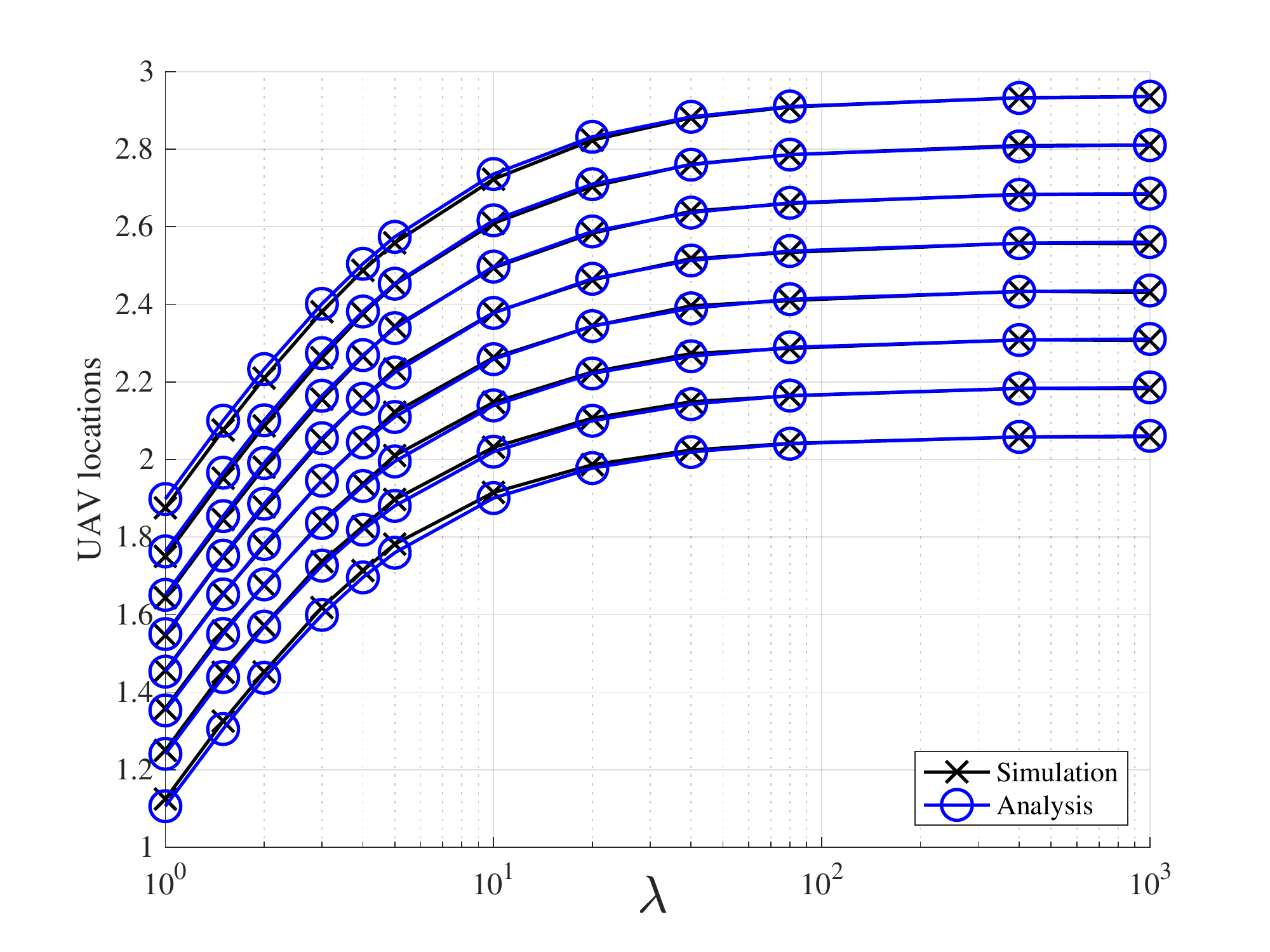}}
\end{center}
\caption{Optimal locations of $n=8$ UAVs with centralized selection.}
\label{fig1}
\end{figure}

\section{Numerical Results}
\label{secNumerical}

In this section, we provide numerical simulation results that verify our analysis. We consider the same setup as in Example \ref{example1}. We use the Lloyd algorithm to design the UAV deployments and compare their performance with our analytical results in Theorems \ref{theorem1} and \ref{theorem2}. In Fig. \ref{fig1}, we show the simulated and analytical optimal UAV locations for the centralized selection scheme and different values of $\lambda$. The analytical results are determined via inverse transform sampling applied to the optimal point density function $\ell(z)$, as in \cite{koyuncu1}. That is to say, given $n$ UAVs, the location of UAV $i$ is estimated as $\mathcal{L}^{-1}(\frac{2i-1}{2n})$, where $\mathcal{L}(z) = \int_0^z \ell(z)\mathrm{d}z$ is the cumulative optimal point density function. This generates a set of points that are faithful to the point density function. In general, we can observe that the simulations match the analysis almost perfectly. When $\lambda = 1$, according to our analysis in Section \ref{secCentralized}, the optimal UAV locations should be designed according to density $Z = \frac{X+Y}{2}$ which is a ``triangular density'' on $[1,2]$. We thus expect the UAVs to be gathered on $[1,2]$ with more concentration around $1.5$ due to the triangle's peak. This is precisely what we observe in Fig. \ref{fig1}. Also, as $\lambda \rightarrow\infty$, we can observe the UAVs tend to a uniformly-spaced configuration over the interval $[2,3]$. This is because as $\lambda \rightarrow \infty$, we have $Z \rightarrow Y$, and since $Y$ is uniform on $[2,3]$, the optimal UAV locations should follow a uniform quantizer structure on $[2,3]$.

\begin{figure}
\begin{center}
\scalebox{0.46}{\includegraphics{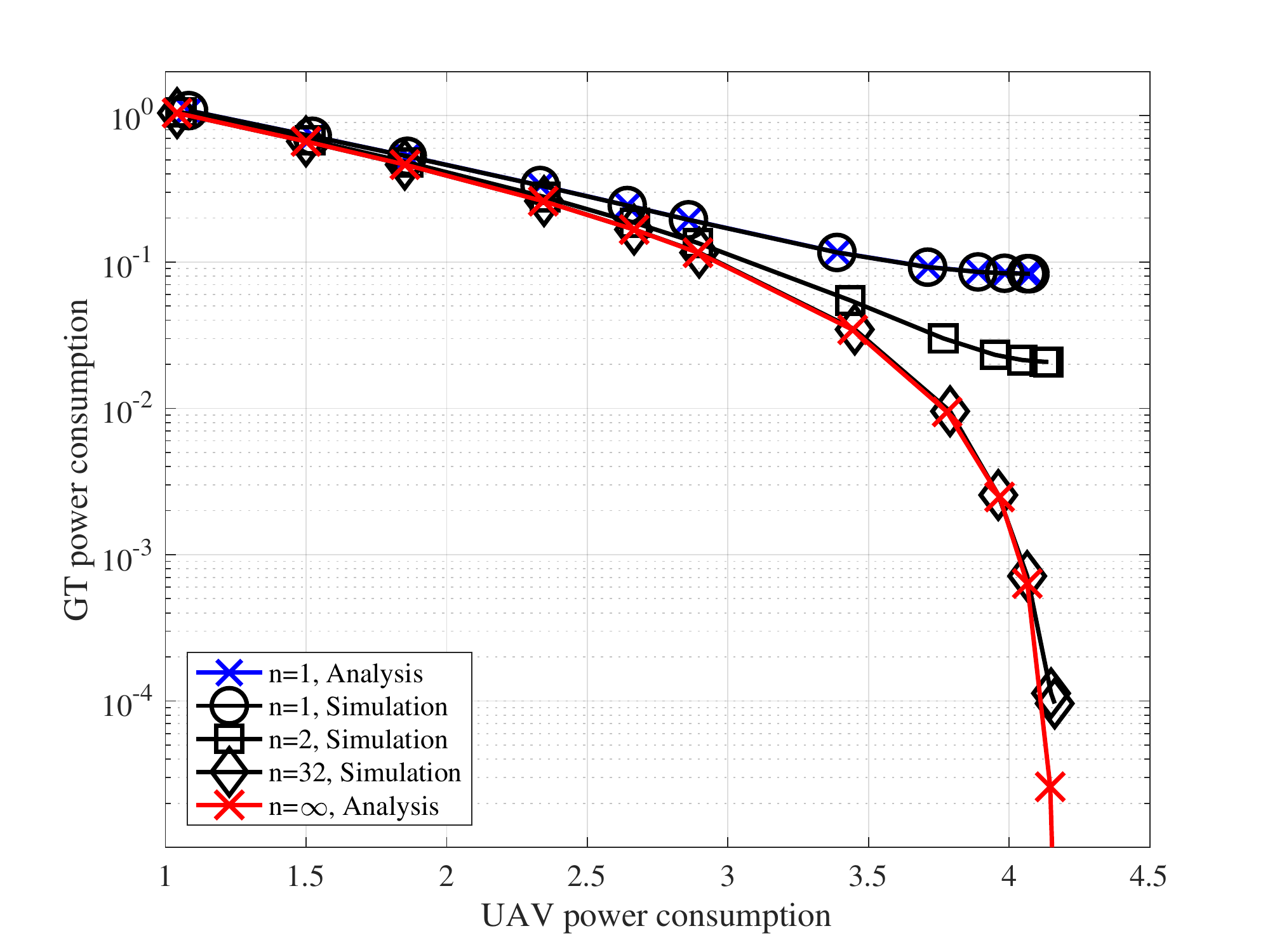}}
\end{center}
\caption{The UAV-GT power consumption tradeoff for centralized selection.}
\label{fig2}
\end{figure}

In Fig. \ref{fig2}, we show the UAV-GT power consumption tradeoff curves with centralized selection. For the particular setup of Example \ref{example1}, if a GT-UAV power consumption pair $(\puav^1,\pgt^1)$ is achievable, then so does $(\pgt^1,\puav^1)$. We thus only show the domain where $\puav^1 \geq 1$ and $\pgt^1 \leq 1$. For $n=1$, we can observe that the analytical results in Example \ref{example1} (which themselves follow from Proposition \ref{oneuavprop}) matches the simulation perfectly. For a general $n$, the GT power consumption saturates to $\frac{1}{12n^2}$ as we allow larger and larger UAV power consumptions. As $n\rightarrow\infty$, the tradeoff  converges to the formula provided by Theorem \ref{theorem1} in Example \ref{example2}, verifying our analysis.

\begin{figure}
\begin{center}
\scalebox{0.46}{\includegraphics{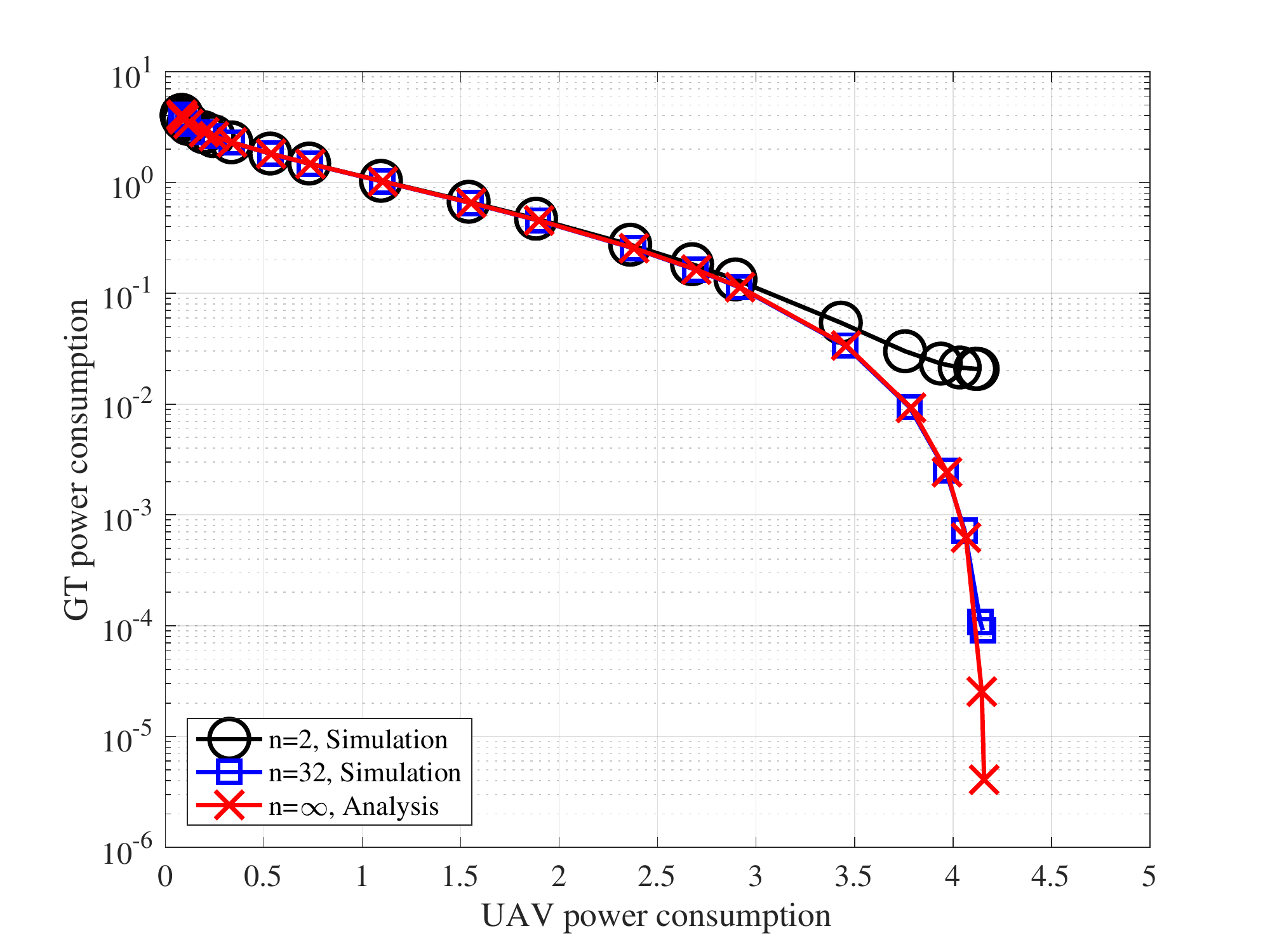}}
\end{center}
\caption{The UAV-GT power consumption tradeoff for distributed selection.}
\label{fig3}
\end{figure}

In Fig. \ref{fig3}, we show the tradeoff curves for distributed selection. The case $n=1$ is identical to that of centralized selection and is thus not shown. Note that the tradeoff is no longer symmetric. In fact, allowing an arbitrarily large GT power consumption does not improve the UAV power consumption beyond $\frac{1}{12}$. This is because, as $\lambda \rightarrow\infty$, we have $W \rightarrow 2.5$. One can thus place all UAVs on $2.5$ without loss of optimality, which achieves a UAV power consumption of $\frac{1}{12}$. Also, the behavior of the case of large UAV power consumption is very similar to what we have observed Fig. \ref{fig1}.  This can also be verified through the  formulae in Example \ref{example2}.

\section{Conclusions}
\label{secConclusions}
We have studied the optimal deployment of UAVs acting as relays between several GTs and GRs for a path loss exponent of $2$. Future work will focus on extensions to a general exponent. Extensions to fading channels, and UAV trajectory optimization for time-varying GT and GR densities are other interesting directions for further research.

\end{document}